\documentclass[aps,prd,showpacs,amssymb,superscriptaddress, notitlepage,floats,floatfix,nofootinbib, twocolumn]{revtex4-2}

\usepackage{graphicx}
\usepackage{dcolumn}
\usepackage{bm}
\usepackage{amsmath}
\usepackage{color}
\usepackage{hyperref}
\usepackage{subfigure}

\begin{document}

\title{Analysis cosmological tachyon and fermion model and observation data constraints
}

\author{P. Tsyba}
\email{pyotrtsyba@gmail.com}
\author{O. Razina}
\email{olvikraz@gmail.com}

\affiliation{Department of General $\&$ Theoretical Physics, \\ L.N. Gumilyov Eurasian National University, Nur-Sultan, 010008, Kazakhstan}

\author{N. Suikimbayeva}
\email{nurgul-1708@mail.ru}
\affiliation{Department of General $\&$ Theoretical Physics, \\ L.N. Gumilyov Eurasian National University, Nur-Sultan, 010008, Kazakhstan}
\affiliation{Department of Physics, \\ M.Kh.Dulaty Taraz Regional University, Taraz, 080012, Kazakhstan}

\date{\today}

\date{\today}

\begin{abstract}
In this work we investigate a cosmological model with the tachyon and fermion fields with barotropic  equation of state, where pressure $p$, energy density $\rho$ and barotropic index $\gamma$ are related by the relation $p=(\gamma-1)\rho$. We applied the tachyonozation method  which  allows to consider cosmological model with the fermion and the tachyon fields, driven by special potential. In this paper, tachyonization model was defined from the stability analysis and exact solution standard of the tachyon field. Analysis of the solution via statefinder parameters illustrated that our model in fiducial points with deceleration parameter $q = 0.5$ and statefinder $r = 1$ which corresponds to the matter dominated universe (SCDM) but, ends its evolution at a point  in the future   $(q =-1, \ r = 1)$ which corresponds to the de-Sitter expansion. Comparison of the model parameters with the cosmological observation data demonstrate, that our proposed cosmological model is stable at barotropic index $\gamma_0=0.00744$.
\end{abstract}

\maketitle

\section{Introduction}

Tachyon is a non-stable field, which might be a basis for modifications of string theory. A tachyon field can play one of the leading roles in the modern accelerated expansion of the universe due to dark energy \cite{Padmanabhan, Feinstein, Abramo, Aguirregabiria1, Aguirregabiria2, Calcagni, Chimento}. Also, it makes a big contribution to inflationary models \cite{Campuzano, Chattopadhyay, Campo, Jain2011} and brings a useful role in cosmology with different types of the tachyon potentials \cite{Bagla, Guo, Copeland, Rudinei, Ivan}.

The tachyon Lagrangian was  extended in the work \cite{Chimento1}, where as a result barotropic index could take the different values, giving rise to phantom and complementary tachyons \cite{Aguirregabiria2, Chimento, Chimento1}. The standard tachyon can behave as an inflaton field in early epoch, and like matter in late epoch.The phantom tachyon field provides rapid expansion and its energy density is gradually increasing. The complementary tachyon field always plays the role of matter \cite{Chimento, Shang-Gang, Rangdee, Bohdan}. In the work \cite{Ivan} was illustrated, that a standard tachyon field, after applying a form invariance symmetry to it, can generate a complementary tachyon field and a phantom tachyon field. In \cite{Chimento} tachyonization of $\Lambda CDM$ model was made for spatially flat, homogeneous and isotropic Friedman-Robertson-Walker (FRW) space-time. There was shown that the standard and complementary tachyon fields in the limiting case generate $\Lambda CDM$ model. The energy-momentum tensor of the tachyon field is the sum of the two tensors associated with the dark matter and the energy density of the vacuum \cite{Chimento2}. The tachyon potential has the non-stable maximum at the origin, decreasing almost to zero as the field grows to infinity.

At present, there are many models of dark energy and any of them can be used to describe the dynamics of the universe \cite{Nojiri, Bamba2, Elizalde2019, Elizalde2018, Elizalde, Cognola2008, Cognola2006, Elizalde2004}. Some of these models are based on scalar fields and some of them are similar to inflationary models \cite{Armendariz-Picon, Armendariz-Picon1, Chiba, De Putter} or based on fermionic fields \cite{Kulnazarov, MyrzakulovY, Razina3, RibasD, Samojeden, Samojeden1, Cai, Wang, Ribas, Yerzhanov, Momeni, Rakhi}.
Particularly, it was shown, that the fermion field plays a crucial role in the isotropization of the initially anisotropic space-time, the formation of a free singularity of cosmological solutions, and the explanation of the late acceleration time.
Models with fields of electrodynamic origin \cite{Razina, Razina1, Razina2} are also of great interest. Action with of the $F_{\mu\nu}$ electromagnetic field tensor or
$F^a_{\mu\nu}$ Yang-Mills fields allows to find new interesting solutions to cosmological problems. Yang-Mills fields can interact with themselves and with each other, that is why they are non-linearity, the superposition principle is not fulfilled for them, and their non-linearity complicates the search for solutions.

For the first time confirmation of the accelerating expansion of the universe appeared in Type Ia supernova observations \cite{Perlmutter,Riess}. Follow-up observations such as: baryon acoustic oscillations \cite{Eisinsten}, large-scale structure of the universe \cite{Tegmark,Seljak}, CMB observation \cite{Spergel,Komatsu}, weak gravitational lensing \cite{Jain}, as well as the calculation of the Hubble parameter depending on the redshift confirmed this phenomenon.

Due to the periodic updating of observational data and the emergence of many theoretical models of dark energy there was a need to create certain statistics, which could differentiate from each other and from the models with a cosmological constant, the models are affected by various types of dark energy \cite{Alam}. One of these statistics is proposed in \cite{Sahni} pair of statefinder parameters $\left\{r, s\right\}$. Statefinder parameters investigate the dynamics of the expansion of the universe through higher-order derivatives of the scale factor. For the $\Lambda CDM$ model in the FRW universe, the fixed point is the value of the parameters of the state determinants $\{r, s \}=\{1, 0 \}$. In our work, we will check if there is a deviation of the model under study from this fixed point.

Above there is a set of works in which models with the tachyon field are investigated. In this paper we consider a model containing the tachyon and the fermionic fields with a barotropic equation of state. A feature of this model is that the addition of the fermionic field imposes additional restrictions on the parameters of the model, and leads to a restriction on the type of the potential of the fermionic field. This shows the possibility of this model. On the other hand, this model makes it possible to cover the dynamics of a wider spectrum of particles in comparison with models containing only the tachyon fields. The investigation model will be tested on Supernovae SN Ia, Hubble, BAO, Planck. Using the observational data, we obtain the value of the barotropic index $\gamma_0$, with the help of which we check the model for stability from the condition for the speed of sound.

The work is organized as follows: in section 2, model with a tachyon field is considered and dynamic equations are constructed; in section 3, we solve the dynamical equations; in section 4, the solution is analyzed by the method of finding state determinants; in section 5, we make use of the observational data to analyze the solution; in section 6, the results of the work are given.

\section{Model}

Let us present of the investigating model described by following Einstein-Hilbert action
\begin{equation} \label{deis}
	S=\frac{1}{16\pi G}\int d^4x \sqrt{-g}\{\frac{1}{2}R+\mathcal{L}_{\phi}+	\mathcal{L}_{D}\},
\end{equation}
where $R$ is the scalar curvature.

The density of the Lagrangian of the tachyon field $\phi$ takes the form 
\begin{equation} \label{}
	\mathcal{L}_{\phi}=-V_1(\phi)\sqrt{1-\partial_{\mu}\phi\partial^{\mu}\phi},
\end{equation}
here $V_1(\phi)$ is the potential of the tachyon field.

The dynamics of the spinor field $ {\psi} $ is given by the density of the Lagrangian
\begin{equation} \label{}
	\mathcal{L}_{D}=\frac{i}{2}[\bar{\psi}\Gamma^{\mu}D_{\mu}\psi-(D_{\mu}\bar{\psi})\Gamma^{\mu}\psi]-V_2(\bar{\psi}\psi),
\end{equation}
where $V_2(\bar{\psi}\psi)$ denotes the self-interacting potential of the spinor field, depending on the bilinear function  $\bar{\psi}\psi$.

By varying the action \eqref{deis} with respect to tetrads the Einstein equations read
\begin{equation} \label{sled}
	R_{\mu\nu}-\frac{1}{2}g_{\mu\nu}R=-T_{\mu\nu},
\end{equation}
where $T_{\mu\nu}$ is the energy-momentum tensor and is equal to
\begin{eqnarray} \label{TeI} \nonumber 
	T_{\mu\nu}&=&\frac{i}{4}\left[\bar{\psi}\Gamma_{\mu}D_{\nu}\psi+\bar{\psi}\Gamma_{\nu}D_{\mu}\psi-D_{\nu}\bar{\psi}\Gamma_{\mu}\psi-D_{\mu}\bar{\psi}\Gamma_{\nu}\psi\right]
	+\\&+&\frac{V_1(\phi)}{\sqrt{1-\nabla_{\rho}\phi\nabla^{\rho}\phi}}\nabla_{\mu}\phi\nabla_{\nu}\phi-  \\
	&-&g_{\mu\nu}\Bigl(\frac{i}{2}[\bar{\psi}\Gamma^{\rho}D_{\rho}\psi-(D_{\rho}\bar{\psi})\Gamma^{\rho}\psi]\nonumber\\
	&-&V_2(\bar{\psi}\psi)-V_1(\phi)\sqrt{1-\partial_{\rho}\phi\partial^{\rho}\phi}\Bigr).\nonumber  
\end{eqnarray}
Variation of the action \eqref{deis} with respect to the tachyon field   $\phi$ the Klein-Gordon equation for tachyon gives
\begin{eqnarray} \label{}
	\nabla^{\mu}\nabla_{\mu}\phi+\frac{1}{2} \frac{\nabla_{\mu}\phi\nabla^{\mu}(\nabla_{\nu}\phi\nabla^{\nu}\phi)}{(1-\nabla^{\nu}\phi\nabla_{\nu}\phi)}+\frac{V_{1\phi}}{V_1}=0.  
\end{eqnarray}

Finally, varying the action \eqref{deis} with respect to  spinor field $\psi$ and its conjugate $\bar{\psi}$ the Dirac equations we obtain

\begin{eqnarray} \label{}
	i\Gamma^{\mu}D_{\mu}\psi-\frac{\partial V_2(\bar{\psi}\psi)}{\partial \bar{\psi}}=0,\\
	iD_{\mu}\bar{\psi}\Gamma^{\mu}+\frac{\partial V_2(\bar{\psi}\psi)}{\partial \psi}=0.
\end{eqnarray}

We assume a spatially flat FRW space-time with metric
	\begin{equation}\label{frw}
ds^2=-dt^2+a(t)^2(dx^2+dy^2+dz^2),
\end{equation}
where $a(t)$ is scale factor of the universe.

Action \eqref{deis} together with metric \eqref{frw} can be written as 
\begin{eqnarray}\label{fM1}
	S=\frac{1}{8\pi G}\int dt \{-3\dot{a}^2a+a^3\frac{i}{2}[\bar{\psi}\gamma^0\dot{\psi}-\dot{\bar{\psi}}\gamma^0\psi]\\
	-a^3V_2-a^3V_1\sqrt{1-\dot{\phi}^2}\}.\nonumber
\end{eqnarray}
We use a homogeneous and isotropic the universe, in which the fields of fermions and tachyons are exclusively functions of time, and in this case the equations of motion for these fields can be written in the form
\begin{equation} \label{dot psi}
\dot{\psi}+\frac{3}{2}H{\psi}+iV_{2u}\gamma^0{\psi}=0,
\end{equation}
\begin{equation} \label{bar dot psi}
\dot{\bar{\psi}}+\frac{3}{2}H\bar{\psi}-iV_{2u}\bar{\psi}\gamma^0=0,
\end{equation}
\begin{equation} \label{Klein-Gordon}
\frac{\ddot{\phi}}{1-\dot{\phi}^2}+3H\dot{\phi}+\frac{V_{1\phi}}{V_1}=0,
\end{equation}
where $u=\bar\psi \psi$ is the bilinear function and $V_{2u}=\frac{dV_2}{du}$. For further calculations, we need an expression following from the definition of the bilinear function $u$ and the equations \eqref{dot psi}-\eqref{bar dot psi}, which is valid for an arbitrary potential $V_2$

\begin{equation}\label{dot u}
\dot u= \dot{\bar\psi} \psi+\bar\psi\dot{\psi}= -3Hu.
\end{equation}

From the Einstein equations \eqref {sled} together with the expression for the energy-momentum tensor \eqref {TeI} we obtain the Friedman equations
\begin{equation} \label{F1}
3H^2=\rho,
\end{equation}
\begin{equation} \label{F2}
3H^2+2\dot{H}=-p,
\end{equation}
where $H =\frac{\dot{a}}{a}$ is the Hubble parameter and the total energy density $\rho$ and the total pressure $p$ are determined by the expressions
\begin{equation} \label{totalrho}
\rho=\frac{V_1}{\sqrt{1-\dot{\phi^2}}}+V_2,
\end{equation}
\begin{equation} \label{totalp}
p=-V_1\sqrt{1-\dot{\phi^2}}-V_2+V_{2u}u.
\end{equation}

We introduce an expression relating the pressure and energy density of the tachyon field by the barotropic index $\gamma$, so that $p=(\gamma-1)\rho$, for analyzing the stability of solutions. In this case, it follows from the equations \eqref{totalrho}, \eqref{totalp} that $\dot{\phi}^2=\gamma$ for $V_2=V_{20} u^{\gamma}$, where $V_{20}$ is constant. Then from the Friedman equations \eqref{F1}-\eqref{F2} follows that
\begin{equation} \label{F11}
3H^2=\frac{V_1}{\sqrt{1-\gamma}}+V_{20}u^{\gamma},
\end{equation}
\begin{equation} \label{F21}
\dot{\rho}+\frac{3\gamma H V_1}{\sqrt{1-\gamma}}+3 V_{20}\gamma Hu^{\gamma}=0.
\end{equation}
Let us differentiate the equation \eqref{F11} with respect to time $t$
\begin{equation} \label{}
\dot{\rho}=\frac{\dot{V_1}}{\sqrt{1-\gamma}}+\frac{\dot{\gamma}V_1}{\sqrt{1-\gamma}(1-\gamma)}+ V_{20}\gamma u^{\gamma-1}\dot{u}.
\end{equation}
Using the equation \eqref{F21} and taking into account \eqref{dot u}, we obtain a differential equation for the barotropic index $\gamma$, in which the function of the fermionic field potential is 
\begin{equation} \label{gamma}
\dot{\gamma}=2(\gamma-1)\left(3H\gamma  +\frac{\dot{V_1}}{V_1}\right).
\end{equation}

To obtain asymptotically stable solutions, the barotropic index should tend to a constant value, namely $ \gamma = \gamma_0 $. In this case, an asymptotic differential equation can be obtained from \eqref{gamma}. For $\gamma=\gamma_0 =const$ it follows that $\dot{\gamma}=0$ and 
\begin{equation} \label{}
(\gamma_0-1)\left(3H\gamma_0 +\frac{\dot{V_1}}{V_1}\right)=0.
\end{equation}
We get $3H\gamma_0+\frac{\dot{V_1}}{V_1}=0$ for $\gamma_0\neq 1$. Solving this differential equation, we get
\begin{equation} \label{V1}
V_1=\frac{V_{10}}{a^{3\gamma_0}}.
\end{equation}

Rewriting \eqref{gamma} using \eqref{V1}  we obtain the following
\begin{equation} \label{gamma1}
\dot{\gamma}=6H(\gamma-1)(\gamma  -\gamma_0).
\end{equation}
The solution of the differential equation \eqref{gamma1} for $\gamma_0 \neq 1 $ is
\begin{equation} \label{gamma11}
\gamma=\frac{\gamma_0-C a^{6(\gamma_0-1)}}{1-C a^{6(\gamma_0-1)}},
\end{equation}
where $C$ is the integration constant.

\section{Solution}
Relation between the energy density of the tachyon field and the scale factor obtained from \eqref{totalrho} and \eqref{V1} is 

\begin{equation} \label{rhophi}
\rho_{\phi}=\frac{V_{10}}{\sqrt{1-\gamma_0}}a^{-3\gamma_0},
\end{equation}
and similar to the ratio of an ideal fluid with a constant barotropic index. Let us find and investigate the exact solution for the tachyon field with the following potential
\begin{equation} \label{V11}
V_1=\frac{\sqrt{1-\gamma_0}}{\gamma_0 \phi^2}.
\end{equation}
The potential diverges at the early time period when $\phi \rightarrow 0$, which corresponds to the behaviour of a typical potential in the context of bosonic string theory. The potential has a unique local maximum at an early epoch and a unique global minimum at a later time, in which $V$ tends to zero \cite{Chimento1}. The global minimum lies at infinity \cite{Shang-Gang}. For this potential, the equation of the tachyon field \eqref{Klein-Gordon} takes the form

\begin{equation} \label{Klein-Gordon1}
\frac{\ddot{\phi}}{1-\dot{\phi}^2}+3H\dot{\phi}-\frac{2}{\phi}=0.
\end{equation}
Several works have found the exact solutions for the tachyon field \cite {Rangdee, Bohdan}. Generalizing the ideas obtained in these works, we study the equations of motion by specifying the linear dependence of the tachyon field on the cosmological time
\begin{equation} \label{phi}
\phi=\sqrt{\gamma_0}t,
\end{equation}
which corresponds to the above expression $\dot{\phi}^2=\gamma$, for $\gamma=\gamma_0$ and $0<\gamma_0 <1$. From the equations \eqref{Klein-Gordon1} and \eqref{phi} we find the Hubble parameter
\begin{equation} \label{habl}
H=\frac{2}{3\gamma_0t}
\end{equation}
and scale factor
\begin{equation} \label{sf}
a=a_0 t^{\frac{2}{3\gamma_0}},
\end{equation}
where $a_0$ is an integration constant equal to $a_0=\left(\frac{\gamma_0^2 V_{10}}{\sqrt{1-\gamma_0}}\right)^{\frac{1}{3\gamma_0}}$.
From the Dirac equations \eqref{dot psi}-\eqref{bar dot psi} taking into account the scale factor \eqref{sf}, we find the spinor field function $\psi$ and the bilinear function $u=\bar{\psi}\psi $

\begin{eqnarray}
  \psi_l&=&\frac{c_l}{t^{\frac{1}{\gamma_o}}}e^{iD(t)}, \ (l=1,2), \\ 
\psi_k&=&\frac{c_k}{t^{\frac{1}{\gamma_o}}}e^{-iD(t)}, \ (k=3,4), \\ 
u&=&\dfrac{c}{a^3}=\dfrac{c}{a_0^3 t^{\frac{2}{\gamma_0}}},
 \end{eqnarray}
where $ c_j $ obeys the following condition $c=|c_{1}|^2+|c_{2}|^2-|c_{3}|^2-|c_{4}|^2=\left(\frac{1}{3\sqrt{1-\gamma_0}}\right)^{\frac{1}{\gamma_0}}$ and
\begin{equation}
D(t)=\frac{V_{20}\gamma^2_0}{\gamma_0-2}\left(\frac{c}{a^3_0}\right)^{\gamma_0-1}t^{\frac{2-\gamma_0}{\gamma_0}}+D_{0j}.
  \end{equation}
The energy density and pressure in general form and componentwise for the tachyon and fermionic fields will take the form, respectively	
\begin{equation}
\rho=\rho_{\phi}+\rho_{f}=\frac{4}{3\gamma_0^2 t^2}, \ \rho_{\phi}=\frac{1}{\gamma_0^2 t^2}, \ \rho_{f}=\frac{1}{3\gamma_0^2 t^2},
  \end{equation}
\begin{equation}
p=p_{\phi}+p_{f}=\frac{4(\gamma_0-1)}{3\gamma_0^2 t^2}, \ p_{\phi}=\frac{\gamma_0-1}{\gamma_0^2 t^2}, \ p_{f}=\frac{\gamma_0-1}{3\gamma_0^2 t^2}.
  \end{equation}
	\section{Statefinder parameters}
	
	The various properties of dark energy are highly dependent on the chosen model. In order to distinguish between different and competing cosmological models involving dark energy, certain evaluation criteria is needed. In the works, \cite{Alam, Sahni} two parameters, which called statefinder are introduced. It makes possible to distinguish several models of dark energy. The $r$ parameter is the next cosmological parameter after the Hubble parameter $H$ and the deceleration parameter $q$, and $s$ is a linear combination of $q$ and $r$ is chosen so that it does not depend on the dark energy density. The statefinder parameters are calculated for different investigated models of dark energy, with a constant and variable parameter of the equation of state $\omega$. In the case of a cosmological constant, the pair $r$ and $s$ takes on a particularly simple form. These parameters contain the scale factor $a(t)$ and its third time derivative $t$
	\begin{equation} \label{sr}
r=\frac{\dddot{a}}{aH^3}=\frac{\ddot{H}}{H^3}-3q-2,
  \end{equation}
\begin{equation} \label{ss}
s=\frac{r-1}{3(q-1/2)},
  \end{equation}
where $q$ is the deceleration parameter equal to $q=-a\ddot{a}/\dot{a}^2=-\ddot{a}/aH^2$. The statefinder parameters are geometric diagnostics, since they are constructed from the space-time metric \cite{Ivan1}.

For the scale factor \eqref{sf}, the statefinder parameters \eqref{sr} and \eqref{ss} are
\begin{equation} \label{sr1}
r=1-\frac{9}{2}\gamma_0+\frac{9}{2}\gamma_0^2,
  \end{equation}
\begin{equation} \label{ss1}
s=\gamma_0.
  \end{equation}
The analysis makes it possible to distinguish between the simplest of all models - the $\Lambda CDM$-model and the investigated model of dark energy. For the $ \Lambda CDM$ -model, the value of the first statefinder parameter  is equal to $r = 1$, even when the density of matter changes from a large value in the early period ($ \Omega_m \cong 1, t << t_0 $) to a small value in the late evolution period ($ \Omega_m \rightarrow 0, t >> t_0 $). Here$ \{r, s \} = \{1, 0 \}$ - is the fixed point for $ \Lambda CDM$-model \cite{Alam}.

The second statefinder parameter $ s $ has properties that complement the properties of the first $r$. Since $s$ is clearly independent of $\Omega_m$, some of the properties belonging to $r$ are violated in the combined pair of state definition parameters $\{r, s \}$.
 
\begin{figure}[h!]

		\includegraphics[width=0.45\textwidth]{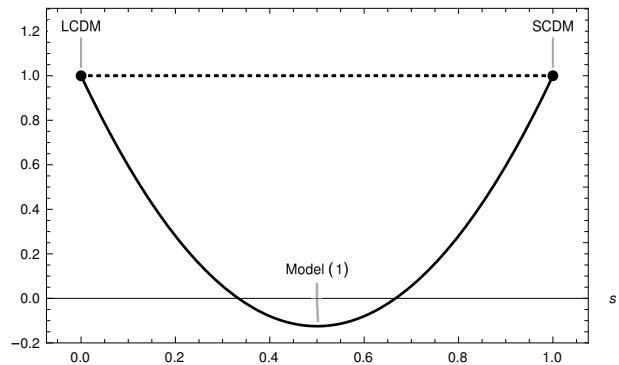}
	\caption{Time evolution of statefinder parameters $\{r, s\}$}
	\label{fig:rs}
\end{figure}

The figure \ref{fig:rs} shows the time evolution of the statefinder parameters of $\{r, s \} $. As can be seen from the figure, $s$ monotonically decreases to zero, and $r$ first decreases from one to a minimum value, and then increases to one. Our model is located to the right of the fixed point of the $\Lambda CDM$ -model ($r = 1, s = 0$).
\begin{figure}[h!]
	\centering
		\includegraphics[width=0.45\textwidth]{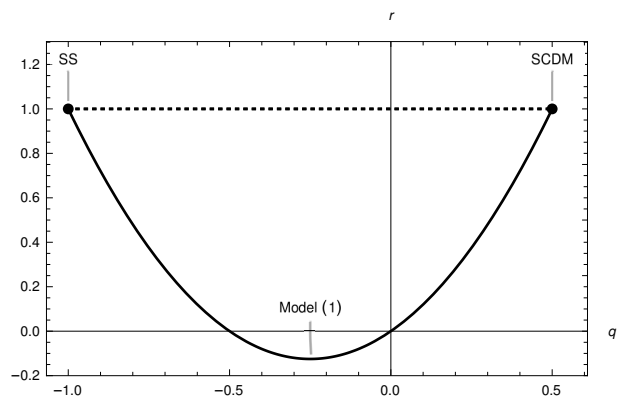}
	\caption{Time evolution of $\{r, q \}$, where $q$ is the deceleration parameter}
	\label{fig:rq}
\end{figure}
Figure \ref{fig:rq} shows the time evolution of $\{r, q \}$, which from the equation \eqref{sr} for the scale factor \eqref{sf} is

\begin{equation} \label{rq}
r(q)=2q^2+q.
  \end{equation}
The dependence graph $\{r, q \}$ passes in the past through the point $(r = 1, \ q = 0.5)$ which corresponds to the matter dominated universe (SCDM) and the point in the future $(r = 1, \ q = - 1)$ which corresponds to steady state model (SS) -- the de Sitter expansion. \textit{The lower the curve on the $s-r$ graph, the less urgent the problem of coincidences becomes. The resulting curve does not qualitatively different from the curves constructed in the previously studied models.
}
\section{Observational Data}
In this work, we use catalogue provided Union 2.1 data which contains 580 points from Type Ia Supernovae \cite{Union21SNe}. BAO data described in Table \ref{TBAO} ~\cite{ Blake12, Busca12, ChuangW12, Chuang13, Anderson13, Anderson14, Font-Ribera13, Delubac14, Percival09, Kazin09, Beutler11, BlakeBAO11, Padmanabhan12, Seo12, Kazin14, Ross14, Aubourg14}. We also use 30 estimations as the Hubble parameters $H(z)$ measured from differential ages of galaxies and BAO data and summarised in Table \ref{TH}, \cite{Simon05, Stern10, Moresco12, Zhang12, Moresco15, Moresco16}. Finally, the CMB parameters are considered from the Plank mission \cite{Planck15}. In order to proceed with the analysis we make use of the technique of the minimum $\chi^2$, which establishes the best set of the parameters.

For comparing the predictions of the model with the above sets of observational data, we use the functions $\chi_i^2(i=SN, H, BAO, CMB)$, and the total function
\begin{equation}
\chi^2_{tot}=\chi^2_{SN}+\chi^2_{H}+\chi^2_{BAO}+\chi^2_{CMB}.\label{chtot}
\end{equation}
	\section{Observational Data}
In this work, we use catalogue provided Union 2.1 data which contains 580 points from Type Ia Supernovae \cite{Union21SNe}. BAO data described in Table \ref{TBAO} ~\cite{ Blake12, Busca12, ChuangW12, Chuang13, Anderson13, Anderson14, Font-Ribera13, Delubac14, Percival09, Kazin09, Beutler11, BlakeBAO11, Padmanabhan12, Seo12, Kazin14, Ross14, Aubourg14}. We also use 30 estimations as the Hubble parameters $H(z)$ measured from differential ages of galaxies and BAO data and summarised in Table \ref{TH}, \cite{Simon05, Stern10, Moresco12, Zhang12, Moresco15, Moresco16}. Finally, the CMB parameters are considered from the Plank mission \cite{Planck15}. In order to proceed with the analysis we make use of the technique of the minimum $\chi^2$, which establishes the best set of the parameters.

For comparing the predictions of the model with the above sets of observational data, we use the functions $\chi_i^2(i=SN, H, BAO, CMB)$, and the total function
\begin{equation}
\chi^2_{tot}=\chi^2_{SN}+\chi^2_{H}+\chi^2_{BAO}+\chi^2_{CMB}.\label{chtot}
\end{equation}
\subsection{Supernovae Ia data}

Union 2.1 compilation consists \cite{Union21SNe} $N_{SN}=580$ SNe Ia with their observed distance module $\mu_i=\mu_i^{obs}$ for redshift $z_i$ in  the interval $0 \leq z_i \leq 1.41$. In order to fit free parameters of our model we compare $\mu_i^{obs}$ with the theoretical value $\mu ^ {th}(z_i)$, where distance modulies is given by
\begin{eqnarray}
\mu (z)\equiv\mu^{th}(z) = 5 \log_{10} \frac{D_L(z)}{10\mbox{pc}}, \\
 D_L (z)= c(1+z) \int_0^z \frac{d\tilde z}{H (\tilde z)} \label{mu}\nonumber.
\end{eqnarray}
Here $D_L (z)$ is the luminosity distance. Corresponding function $\chi^2$ is calculated by taking into account the differences between the SNe Ia observational data and model-specific predictions with parameters $p_1,p_2,\dots$,
 \begin{equation}
\chi^2_{SN}(p_1,p_2,\dots)=\min\limits_{H_0} \sum_{i,j=1}^{N_{SN}}
 \Delta\mu_i\big(C_{SN}^{-1}\big)_{ij} \Delta\mu_j,
  \label{chiSN}\end{equation}
 where $\Delta\mu_i=\mu^{th}(z_i,p_1,\dots)-\mu^{obs}_i$, $C_{SN}$ is the $580\times580$ covariance matrix \cite{Union21SNe}. 

\subsection{BAO data}

Barion acoustic oscillations are obtained from galaxy clustering analysis and include measurements of two cosmological parameters \cite{Eisen05}
\begin{equation}
 d_z(z)= \frac{r_s(z_d)}{D_V(z)},\qquad
  A(z) = \frac{H_0\sqrt{\Omega_m^0}}{cz}D_V(z),
  \label{dzAz} \end{equation}
where $r_s(z_d)$ is the sound horizont at the decoupling epoch and $D_V(z)$ is given by 
$$
D_V(z)=\bigg[\frac{cz D_L^2(z)}{(1+z)^2H(z)}\bigg]^{1/3}.
$$
The values (\ref{dzAz}) was estimated for the redshift $z = z_i$ of galaxies from a peak in the correlation function of the galaxy distribution at the comoving sound horizont scale $r_s(z_d)$, which correspond to the decoupling of the photons $z_d$.  In this work, we use the BAO data from refs. \cite{Blake12, Busca12, ChuangW12, Chuang13, Anderson13, Anderson14, Oka13, Font-Ribera13, Delubac14, Percival09, Kazin09, Beutler11, BlakeBAO11, Padmanabhan12, Seo12, Kazin14, Ross14, Aubourg14} for the parameters (\ref{dzAz}), which provides $ N_ {BAO} = 17 $ data points for $d_z(z)$ and 7 data point for $ A(z)$ and $ C_ {A} $ both are shown in Table  \ref{TBAO}. We use the covariance matrices $C_{d}$ and $C_{A}$ for correlated data from \cite{Percival09, BlakeBAO11}, described in detail in refs. \cite{Sharov21}. So the $\chi^2$ for the values (\ref{dzAz}) yields
\begin{equation}
 \chi^2_{BAO}(p_1,p_2,\dots)=\Delta d\cdot C_d^{-1}(\Delta d)^T+
\Delta { A}\cdot C_A^{-1}(\Delta { A})^T\ ,
  \label{chiB} \end{equation}
 where $\Delta d$ and $\Delta A$ column vector  $\Delta d_i=d_z^{obs}(z_i)-d_z^{th}(z_i)$ and $\Delta A_i=A^{obs}(z_i)-A^{th}(z_i)$.

\begin{table}[ht]
\caption{Values of $d_z(z)=r_s(z_d)/D_V(z)$ and $A(z)$ (\ref{dzAz})
with errors.}
 {\begin{tabular}{||l|l|l|l|l|c|l||}
\hline
 $z$  & $d_z(z)$ &$\sigma_d$    & ${ A}(z)$ & $\sigma_A$  & Refs & Survey\\ \hline
 0.106& 0.336  & 0.015 & 0.526& 0.028& \cite{Beutler11}  & 6dFGS \\ \hline
 0.15 & 0.2232 & 0.0084& -    & -    & \cite{Ross14}& SDSS DR7  \\ \hline
 0.20 & 0.1905 & 0.0061& 0.488& 0.016& \cite{Percival09,BlakeBAO11}  & SDSS DR7 \\ \hline
 0.275& 0.1390 & 0.0037& -    & -    & \cite{Percival09}& SDSS DR7 \\ \hline
 0.278& 0.1394 & 0.0049& -    & -    & \cite{Kazin09}  &SDSS DR7 \\ \hline
 0.314& 0.1239 & 0.0033& -    & -    & \cite{BlakeBAO11}& SDSS LRG \\ \hline
 0.32 & 0.1181 & 0.0026& -    & -    & \cite{Anderson14} &BOSS DR11 \\ \hline
 0.35 & 0.1097 & 0.0036& 0.484& 0.016& \cite{Percival09,BlakeBAO11} &SDSS DR7 \\ \hline
 0.35 & 0.1126 & 0.0022& -    & -    & \cite{Padmanabhan12}   &SDSS DR7 \\ \hline
 0.35 & 0.1161 & 0.0146& -    & -    & \cite{ChuangW12}   &SDSS DR7 \\ \hline
 0.44 & 0.0916 & 0.0071& 0.474& 0.034& \cite{BlakeBAO11}& WiggleZ \\ \hline
 0.57 & 0.0739 & 0.0043& 0.436& 0.017& \cite{Chuang13}& SDSS DR9 \\ \hline
 0.57 & 0.0726 & 0.0014& -    & -    & \cite{Anderson14}& SDSS DR11 \\ \hline
 0.60 & 0.0726 & 0.0034& 0.442& 0.020& \cite{BlakeBAO11} & WiggleZ \\ \hline
 0.73 & 0.0592 & 0.0032& 0.424& 0.021& \cite{BlakeBAO11} &WiggleZ \\ \hline
 2.34 & 0.0320 & 0.0021& -& - & \cite{Delubac14} & BOSS DR11 \\ \hline
 2.36 & 0.0329 & 0.0017& -& - & \cite{Font-Ribera13} & BOSS DR11 \\  \hline
 \end{tabular}
 \label{TBAO}}
\end{table}

\subsection{$H(z)$ data}

The Hubble parameter value $H$ at certain redshift $z$ can be measured with two methods 1) extraction $ H(z)$ from BAO data  \cite{Blake12, Busca12, ChuangW12, Chuang13, Anderson13, Anderson14, Oka13, Font-Ribera13, Delubac14}and 2) making $H(z)$ estimation from differential ages $\Delta t$ of galaxies  \cite{Simon05, Stern10, Moresco12, Zhang12, Moresco15, Moresco16} via the following relation
$$ 
 H (z)= \frac{\dot{a}}{a} \simeq -\frac{1}{1+z}
\frac{\Delta z}{\Delta t}.
 $$
 
 In this paper we used only $N_H=30$ values $H(z)$ estimated from differential ages of galaxies, represented in Table~\ref{TH}. The theoretical values $H^{th}(z_i, p_1,\dots)$ naturally depend on $H_0$. so the $\chi^2$ function is marginalized over $H_0$ \cite{SharovBPNC17}
  \begin{equation}
\tilde\chi^2_{H}= \sum_{i=1}^{N_H} \left[\frac{H^{obs}(z_i)-H^{th}(z_i,
p_j)}{\sigma_{H,i}}\right]^2,\qquad
\chi^2_{H}=\min\limits_{H_0}\tilde\chi^2_{H}.\label{chiH}
\end{equation}

\begin{table}[ht]
\caption{Hubble parameter values $H(z)$ with errors $\sigma_H$ from
refs.~\cite{Simon05,Stern10,Moresco12,Zhang12,Moresco15,Moresco16}} 
 {\begin{tabular}{||l|l|l|c||l|l|l|c||}   \hline
 $z$    & $H(z)$ &$\sigma_H$  & Refs  &   $z$ & $H(z)$ & $\sigma_H$  & Refs\\ \hline
0.070  & 69  & 19.6& \cite{Zhang12}& 0.4783& 80.9& 9   & \cite{Moresco16}\\ \hline
0.090  & 69  & 12  & \cite{Simon05}  & 0.480 & 97  & 62  & \cite{Stern10}    \\ \hline
0.120  & 68.6& 26.2& \cite{Zhang12}& 0.593   & 104   & 13  & \cite{Moresco12} \\ \hline
0.170  & 83  & 8   & \cite{Simon05}&0.6797& 92  & 8   & \cite{Moresco12} \\ \hline
0.1791 & 75  & 4   & \cite{Moresco12}& 0.7812& 105 & 12  & \cite{Moresco12} \\ \hline
0.1993 & 75  & 5   & \cite{Moresco12}& 0.8754& 125 & 17  & \cite{Moresco12}   \\ \hline
0.200  & 72.9& 29.6& \cite{Zhang12}& 0.880 & 90  & 40  & \cite{Stern10}\\ \hline
0.270  &77   &14   &\cite{Simon05} &0.900 & 117 & 23  & \cite{Simon05}  \\ \hline
0.280  & 88.8& 36.6& \cite{Zhang12}& 1.037 & 154 & 20  & \cite{Moresco12}   \\ \hline
0.3519 & 83  & 14  & \cite{Moresco12}& 1.300 & 168 & 17  & \cite{Simon05} \\ \hline
0.3519 & 83  & 14  & \cite{Moresco12}& 1.363 & 160 & 33.6& \cite{Moresco15} \\ \hline
0.3802 & 83  & 13.5& \cite{Moresco16}& 1.430 & 177 & 18  & \cite{Simon05}\\ \hline
0.400 & 95  & 17  & \cite{Simon05}&  1.530 & 140 & 14  & \cite{Simon05} \\ \hline
0.4004& 77  & 10.2& \cite{Moresco16}&  1.750 & 202 & 40  & \cite{Simon05} \\ \hline
0.4247& 87.1& 11.2& \cite{Moresco16}& 1.965 &186.5& 50.4& \cite{Moresco15} \\ \hline
\end{tabular}
  \label{TH}}
 \end{table}

\subsection{CMB data}
\label{CMBdata}

As opposed to the described above SNe Ia, BAO and $H(z)$ data corresponding to the late-time era $0<z\le 2.36$, cosmological observations associated with CMB radiation
\cite {Aubourg14, WangW2013, HuangWW2015} include parameters at the photon-decoupling epoch $z_*\simeq1090$ ($z_*=1089.90 \pm 0.30$ \cite{Planck15}), particularly the comoving sound horizon $r_s(z_*)$  and the transverse comoving distance

\begin{eqnarray}
 r_s(z)&=&\frac1{\sqrt{3}}\int_0^{1/(1+z)}\frac{da}
 {a^2H(a)\sqrt{1+\big[3\Omega_b^0/(4\Omega_r^0)\big]a}}\ , \\
 D_M(z_*)&=&\frac {D_L(z_*)}{1+z_*} = c \int_0^{z_*}
\frac{d\tilde z}{H (\tilde z)}\ . \nonumber
  \label{rs2}\end{eqnarray}

In this work, we use the CMB parameters in the following form \cite{WangW2013,HuangWW2015}

 \begin{equation}
  \mathbf{x}=\big(R,\ell_A,\omega_b\big)=\bigg(\sqrt{\Omega_m^0}\frac{H_0D_M(z_*)}c,\,\frac{\pi
  D_M(z_*)}{r_s(z_*)},\,\Omega_b^0h^2\bigg)
 \label{CMB} 
 \end{equation}
with the estimations (distance priors)\cite{HuangWW2015} 
  \begin{eqnarray}
  R^{Pl}=1.7448\pm0.0054,\\
   \ell_A^{Pl}=301.46\pm0.094,\nonumber \\ 
  \omega_b^{Pl}=0.0224\pm0.00017.\nonumber
   \label{CMBpriors} 
   \end{eqnarray}
   Here $\Omega_b^0$ is the present time baryon fraction. The distance priors (\ref{CMBpriors}) with their errors $\sigma_i$ and the the covariance matrix
   $$C_{CMB}=\|\tilde
C_{ij}\sigma_i\sigma_j\|,\qquad
 \tilde C=\left(\begin{array}{ccc} 1 & 0.53 & -0.73\\ 0.53 & 1 & -0.42\\ -0.73 & -0.42 & 1
\end{array} \right),$$
where derived in ref.\cite{HuangWW2015} from the Planck collaboration data \cite{Planck15} with free amplitude of the lensing power spectrum. For the value $z_*$ we use the fitting formula \cite{WangW2013,HuangWW2015,HuSugiyama95}, the sound horizon $r_s(z_*)$ is estimated from equation \eqref{rs2} as the correction $\Delta r_s=\frac{dr_s}{dz} \Delta z$.
  
     Hence, the $\chi^2$ function corresponding to the data \eqref{CMB} and \eqref{CMBpriors} is obtained as follows
  \begin{eqnarray}
\chi^2_{CMB}=\min_{H_0,\omega_b}\tilde\chi^2_{CMB},\\
\tilde\chi^2_{CMB}=\Delta\mathbf{x}\cdot C_{CMB}^{-1}\big(\Delta\mathbf{x}\big)^{T},\nonumber \\ 
\Delta \mathbf{x}=\mathbf{x}-\mathbf{x}^{Pl} \nonumber
 \label{chiCMB} \end{eqnarray}
 which is minimized by marginalizing over the additional parameter $\omega_b=\Omega_b^0h^2$.
 However, for the joint analysis of $H(z)$ and CMB data, the
marginalization over $H_0$ is calculated simultaneously
  \begin{equation}
\chi^2_H+\chi^2_{CMB}=\min_{H_0}\big(\tilde\chi^2_{H}+\min_{\omega_b}\tilde\chi^2_{CMB}\big).
 \label{chiHCMB} \end{equation}
 
We present the results for the model considered here. The results of calculating the minimum of the function \eqref{chtot} for the  model \eqref{habl} are shown in Fig. \ref{fig:decpyu11}. The figure \ref{fig:decpyu11} shows the corresponding functions of min $ \chi^2_{SNe}$, min $ (\chi^2_{SNe} +\chi^2_{H})$, min $(\chi^2_{SNe}+\chi^2_{H}+\chi^2_{BAO})$ and min $(\chi^2_{SNe}+\chi^2_{H}+\chi^2_{BAO}+\chi^2_{CMB})$. The figure \ref{fig:decpyu12} shows the dependence on min $\chi^2$ on $\gamma_0$ and $\Omega_m$.
\begin{figure}[h!]
\includegraphics[width=0.45\textwidth]{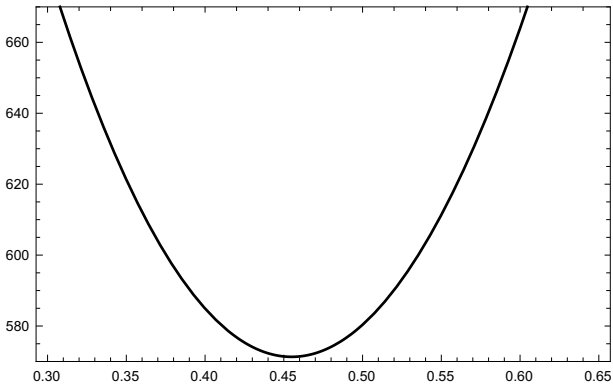}
\caption{Dependence of min$\chi^2$ via $\gamma_0$ for SNe}
\label{fig:decpyu12_1}
\end{figure}
\begin{figure}[h!]
\includegraphics[width=0.45\textwidth]{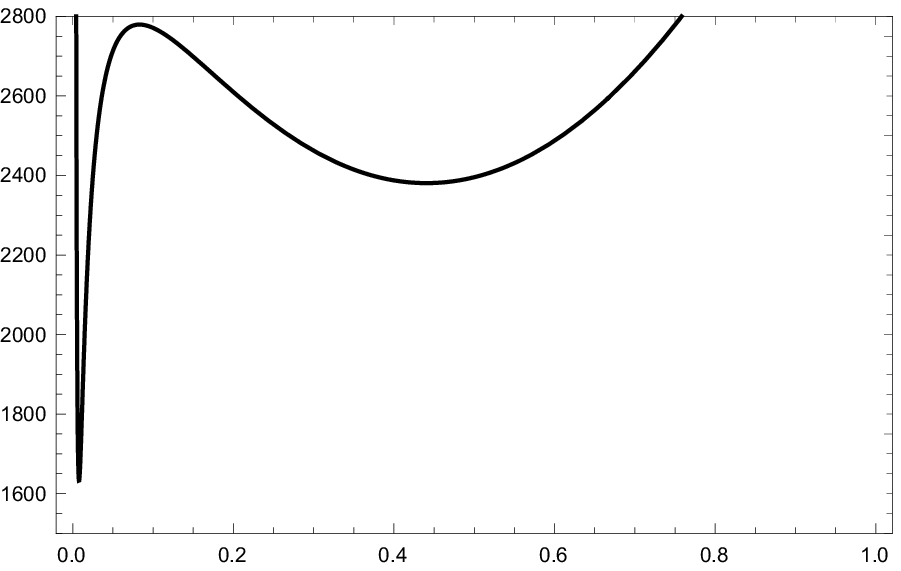}
\caption{Dependence of min$\chi^2$ via $\gamma_0$ for SNe and Hubble}
\label{fig:decpyu12_2}
\end{figure}
\begin{figure}[h!]
\includegraphics[width=0.45\textwidth]{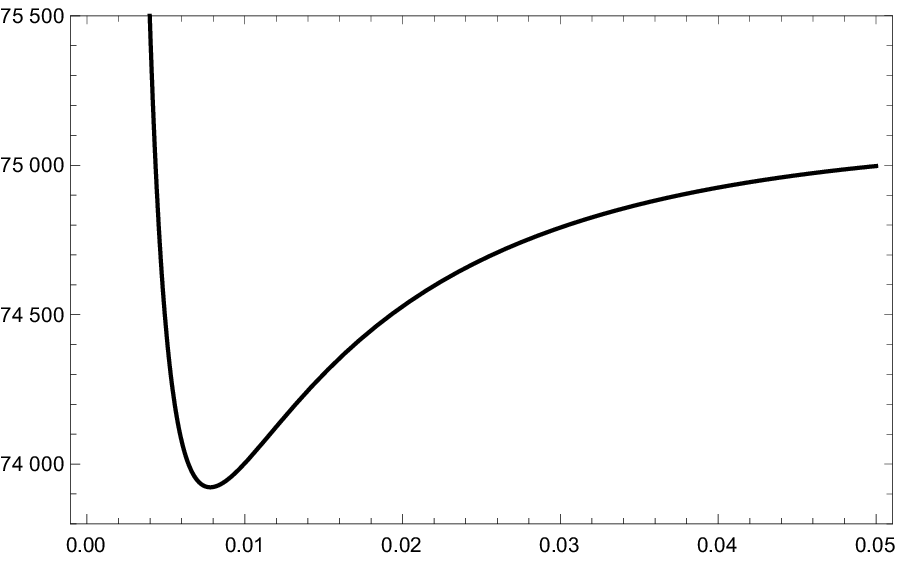}
\caption{Dependence of min$\chi^2$ via $\gamma_0$ for SNe, Hubble and BAO}
\label{fig:decpyu12_3}
\end{figure}
\begin{figure}[h!]
\includegraphics[width=0.45\textwidth]{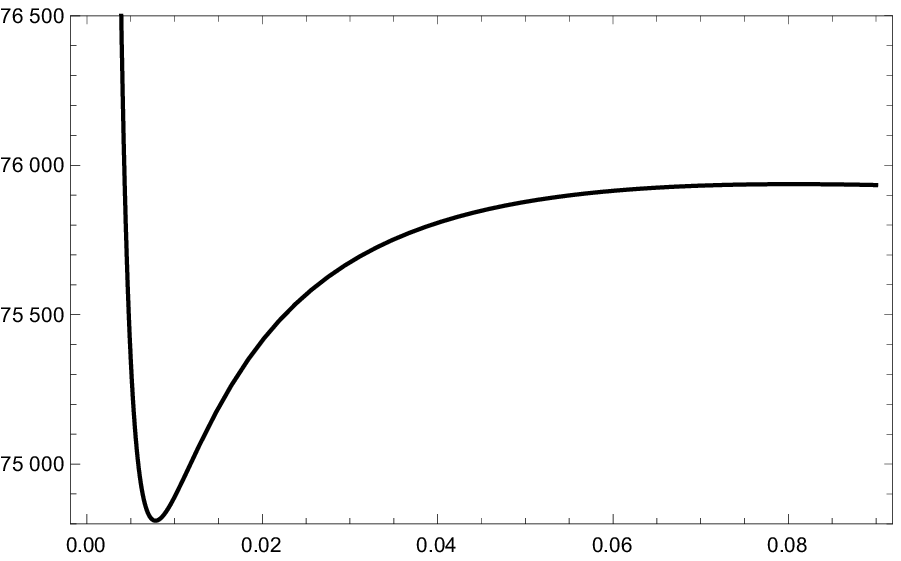}
\caption{Dependence of min$\chi^2$ via $\gamma_0$ for SNe, Hubble, BAO and CMB}
\label{fig:decpyu12_4}
\end{figure}

\begin{figure}[h!]
\includegraphics[width=0.45\textwidth]{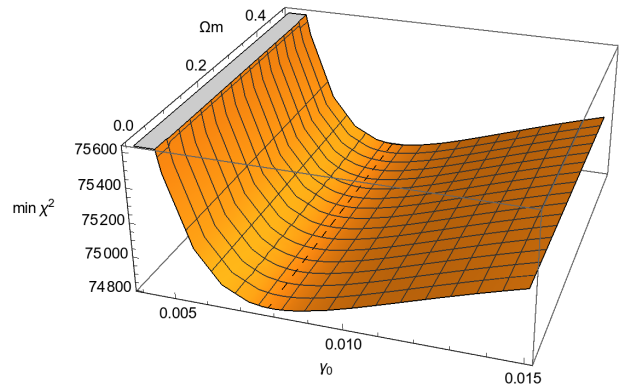}
  \caption{Dependence of min$\chi^2$ via $\gamma_0$ and $\Omega_m$}
\label{fig:decpyu12_5}
\end{figure}

	\section{Conclusion}

The tachyonization method allows us to consider a cosmological model with the fermion field and containing the tachyon field controlled by the potential. We have obtained the standard inverse square potential, which is widely used for tachyons, as well as the corresponding power law solutions for the scale factor. The tachyonization of the model under study was determined from the stability analysis and from the exact solutions of the standard tachyon field controlled by a given potential \eqref{V11}, together with the fermionic field with the controlled potential $V_2=V_{20}u^{\gamma}$.

It is seen that our obtained results, tachyon-fermionic model using statefinder parameters,  are in agreement  with the theory proposed in \cite{Alam}. From the figure \ref{fig:rq} it is clearly seen that our model starts from the predicted point in the past $ (q = 0.5, \ r = 1) $, which corresponds to the SCDM of the universe with a predominance of matter, and ends its evolution at a point in the future $ (q = -1, \ r = 1) $, which corresponds to the de Sitter extension.

Figure \ref{fig:decpyu12_1}, \ref{fig:decpyu12_2}, \ref{fig:decpyu12_3} and \ref{fig:decpyu12_4} demonstrates that as a result of the analysis of the model using supernova SNe we get $\chi^2_{SNe}=571.316 $ and $\gamma_0 = 0.454$ (figure \ref{fig:decpyu12_1}) adding the data Hubble minimum increases $\chi^2_{SNe + H} = 2380.87$, but at the same time the parameter $\gamma_0 = 0.440$ decreases significantly (figure \ref{fig:decpyu12_2}) data from BAO strongly changes the position of the minimum $\chi^2_{SNe + H + BAO} = 73951.6$, but at the same time the parameter $\gamma_0 = 0.0069$ decreases slightly (figure \ref{fig:decpyu12_3}); further application of CMB data leads to a slight increase in $\chi^2_{SNe + H + BAO + CMB} = 74810.6$, and at the same time a slight increase in the parameter $\gamma_0 = 0.00744$ (figure \ref{fig:decpyu12_4}). The large value $\chi^2$ is critical in relation to the Akaike information criterion $AIC = \text{min} \chi^2_{tot} + 2N_p $. However, simultaneously, a decrease in the $\gamma_0$ parameter makes this model stable and provides the possibility of a theory with an ordinary tachyon according to the  criterion \cite{Ivan}. The speed of sound for this model, which is confirmed by the observational data, is $c_s^2 = 1-\gamma_0 = 0.992>1/5. $ Considering the Lagrangian of the tachyon field with allowance for $\dot \phi^2 = \gamma_0 = 0.00744 $, which is small, it can be seen that only the potential part will make the main contribution to the Lagrangian of the tachyon field. That is, the kinetic energy of the tachyon field will be small compared to the potential energy. This leads to a slow roll effect in the late universe, that is, accelerated expansion. From the figure \ref{fig:decpyu12_5} it follows that the model under study has a minimum only with respect to the parameter $\gamma_0$ and the impossibility of achieving minimum with respect to the parameter $\Omega_m$. Along with the obtained advantages, the tachyon-fermionic model has a disadvantage - the Akaike parameter is high compared to the $\Lambda CDM$ model. Another drawback of the investigated model is the difference of the $\Omega_m$ parameter from the predictions of the $\Lambda CDM $-model.

\section*{Acknowledgments}

This study was funded by the Science Committee of the Ministry of Education and Science of the Republic of Kazakhstan AP08955524.


\end{document}